\documentclass[pre,twocolumn,noshowpacs]{revtex4}
\usepackage{epsfig}

\begin{document}

\title{Reply to the comment on ``Non-Normalizable Densities in Strong Anomalous Diffusion:
Beyond the Central Limit Theorem"
}

\author{Adi Rebenshtok}
\affiliation{Department of Physics, Institute of Nanotechnology and Advanced Materials, Bar-Ilan University, Ramat-Gan, 52900, Israel}
\author{Sergey Denisov}
\affiliation{Institute  of Physics, University of Augsburg,
Universit\"atsstrasse 1, D-86135,  Augsburg
Germany}
\affiliation{Sumy State University, Rimsky-Korsakov Street 2, 40007 Sumy, Ukraine}
\affiliation{Department for Bioinformatics, Lobachevsky State University, Gagarin Avenue 23, 603950 Nizhny Novgorod, Russia}
\author{Peter H\"anggi}
\affiliation{Institute  of Physics, University of Augsburg,
Universit\"atsstrasse 1, D-86135,  Augsburg
Germany}
\author{Eli Barkai}
\affiliation{Department of Physics, Institute of Nanotechnology and Advanced Materials, Bar-Ilan University, Ramat-Gan, 52900, Israel}

\begin{abstract}

 We provide a reply to a comment by I. Goychuk 	arXiv:1501.06996 [cond-mat.stat-mech](not under
   active consideration with Phys. Rev. Lett.) 
on  our Letter A. Rebenshtok, S. Denisov, P. H\"anggi, and E. Barkai,
{\em Phys. Rev. Lett.} {\bf 112}, 110601 (2014).

\end{abstract}

\maketitle

 Infinite ergodic theory is a branch of mathematics which  deals with
dynamical systems whose invariant density is infinite,
namely non-normalized densities describe statistical  properties of the system.
 This theory was  applied to
simple deterministic  transformation rules like the Pomeau-Manneville
map,  a well known model for intermittency,
where an infinite  invariant density,
namely a non-normalizable state  describes long time aspects of
the dynamics. In this context mathematicians consider two classes
of observables, integrable and non-integrable
with respect to the infinite density. Only recently has this concept
gained attention in Physics (see Refs. in \cite{PRL}).

 Indeed at a first glance infinite densities (or more specialized infinite
covariant densities)
might seem strange, since a system where the  number of particles is conserved,
must obey rigorous normalization condition for all times as was pointed out in the comment \cite{GoyCom}
and well known from elementary courses.
In our
work we investigated the L\'evy walk model, a widely applicable norm conserving
 model
of super-diffusion, showing that it is described by an infinite
covariant density. Below we show that the comment recently published
on cond-mat,
is based on  misinterpretations of the concepts and results presented
in our work.

To begin with, the author of the comment writes that we claimed that
particle distribution or equivalently probability densities can become
non-normalized in the case of anomalous L\'evy walk. This is a
false accusation. The probability density function of the particle's
position in space, is perfectly normalized, at all times. In our work
we used the concept of infinite covariant densities, i.e. non-normalized densities. However, these densities are not probability densities, as is well
known, and clear from our discussion (see some details below).
It seems that the author of the comment has assigned to the infinite
density a meaning of a probability density, which is wrong.

 To further discuss the comment we first present briefly our main
results. This is required since the author of the comment has not presented
a full picture of our results. We investigated the L\'evy walk model and showed
that in the super diffusive phase,
the center part of the distribution $P_{{\rm cen}}(x,t)$ Eqs. (14,15)
\cite{PRL}
is described by a L\'evy stable law. This distribution is perfectly
normalized, and a standard tool in the field.  In addition we defined
the infinite covariant density Eq. (9)
$$
\lim_{t \to \infty} t^\alpha P(x,t) = I_{\rm cd} (x/t).
$$
Since $P(x,t)$ is the density, which as mentioned is perfectly normalized
$\int_{-\infty} ^\infty P(x,t) {\rm d} x= 1$ and since we have $1<\alpha<2$ it
is not surprising that the infinite density is not normalized,
since the spatial integration over the left hand side of the equation gives
infinity since $t^\alpha \to \infty$ (this conclusion is reached by author of the comment, a trivial insight which is clear from \cite{PRL}).
 It is also very clear,
even without gaining any insight on the subject, that the infinite
covariant density is not a probability density function.
 The goal
of our paper was to show in what sense does the concept of infinite
density describe the statistical properties of the L\'evy walk model,
 and
to obtain analytical expressions for this little understood
function.
First insight was that from data,
e.g. numerical simulations of the model or in principle experiments, one
may  construct a histogram
and then plot $t^\alpha P(x,t)$ versus $x/t$ and then observe that
 in the limit of large time $t$
this scaled histogram  approaches the analytical expressions for the infinite
density given explicitly in \cite{PRL}.
 This is in principle
easy to check numerically, and it is only a pity that the author
of the comment did not find time to do so (i.e., repeat simulations
presented in our work \cite{remark}).
 Secondly the infinite density describes
high order moments, $\langle |x|^q \rangle$ with $q>\alpha$.
In this sense high order moments are integrable with respect to the
infinite density, and their asymptotic values are obtained from
this non-normalized density. In contrast, moments with $q<\alpha$,
including the normalization, $q=0$, are non-integrable with respect
to the infinite density, and hence they are computed with respect
to the L\'evy distribution $P_{{\rm cen}}(x,t)$.
  This in turn is related to strong anomalous
diffusion, a behavior found in many  systems.

 We therefore found it very disturbing to read that according to the
author of the comment, {\em Eq. (8) cannot be applied to the whole range
of $x$ variation}. As mentioned, Eq. (14,15) in \cite{PRL}
explicitly give the behavior of the  center part of the packet of particles
$P_{{\rm cen}}(x,t)$, in terms of symmetric L\'evy distributions.
Thus the  center part of the packet  is described by
a L\'evy distribution, so clearly the  infinite density is not
applicable in the whole range of $x$.
Eq. (18) gives $\langle |x|^0 \rangle=1$, namely the normalization
condition holds as it should. Hence the author's suggestion
 that we claimed that
{\em probability densities can become non-normalizable in the case
of anomalous diffusion} is detached from the reality of our Letter.
 Similarly, the author of the comment also writes
on our  Eq. (13), which is an equation for the moments of the process,
that it  {\em is generally wrong, e.g. for $q<\alpha$}.
However any one reading  our paper sees one line before
Eq.  (13) that it is valid under the explicitly stated condition
that  $q> \alpha$. Our Eq. (18) gives
the solution for $q<\alpha$. Unfortunately we see that
the author of the comment did not present our results decently,
since he does not provide the conditions under which our formulas work,
these being clearly stated in \cite{PRL}.

 Let us turn back to the general philosophy of the comment since
it presents a matter of opinion, namely  that the infinite
density cannot reflect physical reality. First as mentioned
if we scale numerical or experimental data as $t^\alpha P(x,t)$ and
plot it versus $x/t$ the plot will approach the infinite density
with its characteristic non integrable
pole for small argument of $\overline{v}= x/t$ (see figures
in \cite{PRL}). In that sense the infinite density reflects
physical reality. The claim made by  the author of the comment
 is that {\em any descent
experiment, either real or numerical, done at finite time $t$ will
yield $I_{{\rm cd}}(\overline{v},t)$ which is perfectly normalized,
and not $I_{{\rm cd}}(\overline{v})$} and similarly {\em Importantly, stochastic numerics can be done only at finite $t$}. The general claim that
experiments are done on finite time and hence asymptotic results
have no value is a philosophy which is long abandoned. At the starting
days of diffusion theory, one could wonder if the  diffusion equation
is correct. The  solution of the diffusion equation,
predicts  that after a fraction of a second a particle which started
in New York city, can have a finite probability (though small) to be
in Tokyo. Does this imply that we should throw away the diffusion
equation, or for that aim the Gaussian central limit theorem,
which is also valid only in a limit? Clearly asymptotic laws should apply
within their limitations, and the same is true for the infinite
density.

  In the  case of the L\'evy walk and its  infinite
density, which is of-course  an asymptotic result, the situation
is even sharper.
Physics actually {\it begs} for the infinite
density concept.  The center part of the packet is described by
the L\'evy stable law, as mentioned.
It therefore has the awkward property, at first
glance only, that $\langle x^2 \rangle=\infty$ at a finite asymptotic time,
 since the second moment
of a L\'evy distribution is infinite as is well known. Thus, the L\'evy
central limit theorem predicts an unphysical behavior that particles
can travel with an infinite speed.
 Should we reject this limit theorem, because at any finite
time we cannot attain this divergence? In fact the infinite
density concept cures this unphysical behavior. The second moment
$\langle x^2 \rangle$ is finite if one uses the infinite density concept
as we have shown in \cite{PRL}.
We see that the L\'evy walk process has two scaling solutions.
One is the familiar  L\'evy density which gives a finite normalization
and a diverging mean square displacement, the other is the
infinite density which is not normalized but provides the second
moment correctly. The second moment
is  easily considered the most frequently measured moment in diffusion
processes. Both these densities are essential for the correct
statistical description of the process, both can be measured by proper
scaling of data,
both are strictly valid in the long time limit,
and the practitioner in the field will no doubt be able to comprehend
their domain of validity and avoid abuse.

 In summary, the authors cannot be held responsible if others, intentionally
or unintentionally, misrepresent our correct results by not observing the conditions under which those are obtained.

{\bf Acknowledgement} this work was supported by the Israel Science Foundation.

\vspace{2pc}

\end{document}